# A Comparative Study of $\frac{K^{\pm}}{\pi^{\pm}}$ ratio in Proton-Proton Collisions at different energies ---Experimental Results vs. Model Simulation

**Swarnapratim Bhattacharyya[1], Maria Haiduc[2], Alina Tania Neagu[2] and Elena Firu[2]**

[1]**Department of Physics, New Alipore College, L Block, New Alipore, Kolkata 700053, India**

Email: swarna_pratim@yahoo.com

[2]**Institute of Space Science, Bucharest, Romania**

PACS No. 25.75.-q, 25.75.Ag., 25.75.Gz., 24.60.Ky.

**Key Words: Kaon Production, kaon to pion ratio, pp collisions**

## Abstract

A detailed study of energy dependence of $\frac{K^+}{\pi^+}$, $\frac{K^-}{\pi^-}$ and total kaon to pion multiplicity ratio $\left(\frac{K^++K^-}{\pi^++\pi^-} = \frac{K}{\pi}\right)$ has been carried out in proton-proton (pp) collisions at $\sqrt{s}$ = 6.3, 17.3, 62.4, 200 and 900 GeV and also at $\sqrt{s}$ =2.76 TeV and 7 TeV in the framework of UrQMD and DPMJET III model. Dependence of $\frac{K^+}{\pi^+}$ and $\frac{K^-}{\pi^-}$ on energy shows different behavior for UrQMD and DPMJET III model. The presence of the horn like structure in the variation of $\frac{K^+}{\pi^+}$ and $\frac{K^-}{\pi^-}$ with energy for the experimental data is supported by the DPMJET III model. Experimentally it has been observed that as energy increases, the total kaon to pion multiplicity ratio $\left(\frac{K^++K^-}{\pi^++\pi^-} = \frac{K}{\pi}\right)$ increases systematically for pp collisions at lower energies and becomes independent of energy in LHC energy regime. Our analysis on total kaon to pion multiplicity ratio $\left(\frac{K^++K^-}{\pi^++\pi^-} = \frac{K}{\pi}\right)$ with UrQMD data is well supported by the experimental results obtained by different collaborations in different times. In case of DPMJET III data, the saturation of $\frac{K}{\pi}$ ratio at LHC region has not been observed.

## 1. Introduction

The study of nucleus-nucleus interactions at high energies has been a subject of major interest to the theoretical and experimental physicists. The nucleus-nucleus interaction can provide valuable information on the spatiotemporal development of multiparticle production process, which is one of the prime interests in view of recent developments of quantum chromodynamics. Along with the study of nucleus-nucleus collisions, a thorough understanding of proton-proton (pp) collisions is also necessary both as input to detailed theoretical models of strong interactions, and as a baseline for understanding the nucleus–nucleus collisions at relativistic and ultra-relativistic energies. Soft particle production from ultra-relativistic pp collisions is also sensitive to the flavor distribution within the proton, quark hadronization and baryon number transport. The measurement of charged particle transverse momentum spectra in pp collisions serves as a crucial reference for particle spectra in nucleus-nucleus collisions. A proton-proton reference spectrum is needed for nucleus-nucleus collisions to investigate possible initial-state effects in the collision. The multiplicity distribution of particles produced in proton-proton (pp) collisions and the multiplicity dependence of other global event characteristics represent fundamental observables reflecting the properties of the underlying particle production mechanisms. In high energy collisions along with the pions, kaons are also important as the strange particle production is a powerful probe into the hadronic interaction and the hadronization process in pp and heavy ion collisions at relativistic energies. The study of $\frac{K}{\pi}$ ratio in high energy collisions is an important observable to be studied not only to address questions of the phase transition but also to obtain a better understanding of the pre-equilibrium dynamics, the hadronization processes and dynamics of hadrons in the medium. It is well known that the strangeness enhancement in relativistic nucleus-nucleus collisions has been proposed as a signature of the Quark-Gluon Plasma (QGP) formation in the relativistic heavy ion collisions. The study of $\frac{K}{\pi}$ ratio in pp collisions can provide a baseline to investigate the strangeness enhancement.

In this paper, we are presenting an analysis of energy dependence of $\frac{K^+}{\pi^+}$, $\frac{K^-}{\pi^-}$ and total kaon to pion multiplicity ratio $\left(\frac{K^++K^-}{\pi^++\pi^-} = \frac{K}{\pi}\right)$ at $\sqrt{s} = 6.3, 17.3, 62.4, 200$ and $900$ GeV and also at $\sqrt{s}$ =2.76TeV and 7 TeV in the framework of UrQMD and DPMJET III model in proton-proton (pp) collisions. We have also compared our results with available experimental results obtained so far. Before going into the details of the analysis it will be convenient for the readers to have brief introductions about the two models.

## 2. UrQMD and DPMJET III model----A Brief Introduction

UrQMD model is a microscopic transport theory, based on the covariant propagation of all the hadrons on the classical trajectories in combination with stochastic binary scattering, color string formation and resonance decay. It represents a Monte Carlo solution of a large set of coupled partial integro-differential equations for the time evolution of various phase space densities. The main ingredients of the model are the cross sections of binary reactions, the two-body potentials and decay widths of resonances. The UrQMD collision term contains 55 different baryon species (including nucleon, delta and hyperon resonances with masses up to 2.25 GeV/$c^2$) and 32 different meson species (including strange meson resonances), which are supplemented by their corresponding anti-particle and all isospin-projected states. The states can either be produced in string decays, s-channel collisions or resonance decays. This model can be used in the entire available range of energies from the Bevalac region to RHIC. For more details about this model, readers are requested to consult [1-3].

The Monte Carlo event generator DPMJET can be used to study particle production in high-energy nuclear collisions including photo-production and deep inelastic scattering off the nuclei. It is a code system based on the Dual Parton Model and unifies all features of the DTUNUC-2, DPMJET-II and PHOJET1.12 event generators. DPMJET-III allows the simulation of hadron-hadron, hadron-nucleus, nucleus-nucleus, photon-hadron, photon-photon and photon-nucleus interactions from a few GeV up to the highest cosmic ray energies. DPMJET is an implementation of the two-component Dual Parton Model for the description of interactions involving nuclei. This model is based on the Gribov-Glauber [4-6] approach. Gribov theory of high–energy interactions of hadrons and nuclei is based on general properties of amplitudes in relativistic quantum theory and provides a unified approach to a broad class of processes. According to this theory, the Glauber approximation [6] to nuclear dynamics is valid in the region of not too high energies and should be modified at energies of RHIC and LHC. Gribov theory then allows to determine the corrections to the Glauber approximation [6] for inclusive particle spectra by relating them to cross sections of large–mass diffraction. The technique has been applied to calculation of shadowing effects for structure functions of nuclei and a good agreement with experimental data on these processes has been obtained. The same approach predicts a strong reduction of particle densities at super–high energies as compared to predictions of the Glauber approximation [6]. Since its first implementations [7-8] DPMJET model uses the Monte Carlo realization of the Gribov-Glauber multiple scattering formalism according to the algorithms of [9] and allows the calculation of total, elastic, quasi-elastic and production cross sections for any high-energy nuclear collision. DPMJET III is a string model and the generalization of the string model to hadron-nucleus and nucleus-nucleus collisions was done by the Glauber-Gribov theory [10-12]. DPMJET III model treats both soft and hard scattering processes in a unified way. Soft processes are parametrized according to Regge-phenomenology whereas lowest order perturbative QCD is used to simulate the hard component. In DPMJET III model multiple parton interactions in each individual hadron/nucleon/photon-nucleon interaction have been described by the PHOJET event generator and the fragmentation of parton configurations is treated by the Lund model PYTHIA. For more details about the model, one can consult [13-14].

### 3. Analysis & Results

We have generated a large sample of events (Ten thousand events for each case) using the UrQMD (UrQMD-3.3p1) [1-3] and DPMJET III (DPMJET 3.06) [13-14] model in pp collisions at $\sqrt{s}$ = 6.3, 17.3, 62.4, 200 and 900 GeV and also at $\sqrt{s}$ =2.76 TeV and 7 TeV. We have calculated the number of positive and negative kaons and the number of positive and negative pions from the generated output of the UrQMD and DPMJET III model for all the energies.

### 3.1 Energy Dependence Study of $\frac{K^+}{\pi^+}$ ratio

The values of $\frac{K^+}{\pi^+}$ ratio have been calculated from the generated output of both UrQMD and DPMJET III model. Table 1 represents the values of $\frac{K^+}{\pi^+}$ ratio in pp collisions at $\sqrt{s}$ = 6.3, 17.3, 62.4, 200 and 900 GeV and also at $\sqrt{s}$ =2.76TeV and 7 TeV. From Table 1 it is reflected that DPMJET III model simulated values of $\frac{K^+}{\pi^+}$ ratio are higher than their UrQMD counterparts up to $\sqrt{s}$ =62.4GeV. From $\sqrt{s}$ =200GeV, UrQMD simulated values of $\frac{K^+}{\pi^+}$ ratio overestimate the DPMJET III simulated values.

For comparison in table 1 we have shown the experimental values of $\frac{K^+}{\pi^+}$ ratio obtained from different experimental works at $\sqrt{s}$ =6.3 GeV [15], $\sqrt{s}$ =17.3 GeV [16], $\sqrt{s} = 62.4$ GeV [17], $\sqrt{s}$=200GeV [18] and $\sqrt{s}$ =7000GeV [19]. S. Pulawski presented [15] the experimental values of $\frac{K^+}{\pi^+}$ ratio at mid rapidity at $\sqrt{s}$ =6.3 GeV [15] in case of inelastic pp collisions for the data of NA61/SHINE collaboration. According to his study NA61/SHINE data suggests that the energy dependence of $\frac{K^+}{\pi^+}$ ratio exhibits rapid changes in the SPS energy range. S. Pulawski pointed out that [15] the EPOS, UrQMD, Pythia 8 and HSD model failed to describe the NA61/SHINE experimental results satisfactorily. The values of $\frac{K^+}{\pi^+}$ ratio $\sqrt{s} = 62.4$ GeV and 7 TeV have been calculated from the $\frac{dN}{dy}$ values at mid rapidity. The $\frac{dN}{dy}$ values at mid rapidity at $\sqrt{s} = 62.4$ GeV and at $\sqrt{s} = 7$ TeV are |y| < 0.35 and |y| < 0.5 respectively. $\frac{K^+}{\pi^+}$ ratios at other energies have been estimated with the help of a high accuracy digitized plot analyzer.

In figure 1 we have presented the variation of $\frac{K^+}{\pi^+}$ ratio with energy for pp collisions in case of UrQMD simulated, DPMJET III simulated and the experimental values. From the figure it can be seen that for UrQMD simulation the values of $\frac{K^+}{\pi^+}$ ratio increase smoothly with energy and after reaching at $\sqrt{s}$ =900 GeV the ratio almost saturates. While in case of DPMJET III simulation the value of $\frac{K^+}{\pi^+}$ ratio shows a sudden decrease at $\sqrt{s}$ =200 GeV. After the sudden decrease the $\frac{K^+}{\pi^+}$ values begin to increase with energy. No prominent saturation of $\frac{K^+}{\pi^+}$ ratio has been observed at LHC regime for DPMJET III model. The dependence of $\frac{K^+}{\pi^+}$ values with energy shows the presence of the horn like structure in case of DPMJET III simulated events. The observed sudden decrease of $\frac{K^+}{\pi^+}$ ratio is completely absent in case of UrQMD simulation. The experimental studies of energy dependence of $\frac{K^+}{\pi^+}$ ratio also indicate that the $\frac{K^+}{\pi^+}$ values increase initially with energy, get a sudden drop at $\sqrt{s} = 62.4$ GeV and go on increasing again signifying the presence of the horn like structure.

### 3.2 Energy Dependence Study of $\frac{K^-}{\pi^-}$ ratio

In order to study the energy dependence of $\frac{K^-}{\pi^-}$ ratio in pp collisions we have calculated the values of $\frac{K^-}{\pi^-}$ ratio obtained from the simulation of pp collisions at $\sqrt{s}$ =6.3GeV-7TeV by UrQMD and DPMJET III model. Calculated values of $\frac{K^-}{\pi^-}$ ratio for UrQMD and DPMJET III simulation have been presented in table 2. From table 2 it is seen that from $\sqrt{s}$ =900GeV, UrQMD simulated values of $\frac{K^-}{\pi^-}$ ratio overestimate the DPMJET III simulated values. In the same table the values of $\frac{K^-}{\pi^-}$ ratio calculated from the different experimental publications at $\sqrt{s}$ =6.3 GeV [15], $\sqrt{s}$ =17.3 GeV [16], $\sqrt{s} = 62.4$ GeV [17], $\sqrt{s}$=200GeV [18], $\sqrt{s} = 900$ GeV [20] and $\sqrt{s}$ =7000GeV [19] have also been presented. As in the case of $\frac{K^+}{\pi^+}$ ratio, the values of $\frac{K^-}{\pi^-}$ ratio have been calculated from the values of $\frac{dN}{dy}$ at mid rapidity at $\sqrt{s} = 62.4$ GeV (|y| < 0.35 ) and $\sqrt{s} = 7$ TeV (|y| < 0.5).

In figure 2 we have depicted the variation of $\frac{K^-}{\pi^-}$ ratio with energy for UrQMD simulated, DPMJET III simulated and the experimental values. From figure 2 it may be noted that the values of $\frac{K^-}{\pi^-}$ ratio for the UrQMD simulated events are found to increase smoothly with energy and after reaching $\sqrt{s}$=900GeV, saturation of $\frac{K^-}{\pi^-}$ ratio occurs. However in case of DPMJET III model a sudden drop of $\frac{K^-}{\pi^-}$ ratio occurs at $\sqrt{s}$=900GeV. The ratio then begins to rise again presenting a horn like structure as observed in case of the energy dependence of $\frac{K^+}{\pi^+}$ values. The experimental values of $\frac{K^-}{\pi^-}$ ratio also get a sudden drop at $\sqrt{s} = 62.4$ GeV and increase again to construct a horn like structure in the energy dependence of $\frac{K^-}{\pi^-}$ values in pp collisions.

Thus it may be pointed out that the experimental study of energy dependence of both $\frac{K^+}{\pi^+}$ and $\frac{K^-}{\pi^-}$ ratio shows a horn like structure which is also shown by the DPMJET III model but UrQMD model fails to reproduce the horn like structure. The observed difference in the energy dependence of $\frac{K^+}{\pi^+}$ and $\frac{K^-}{\pi^-}$ ratio between UrQMD and DPMJET III model is due to the basic difference between the two models. It may be mentioned here that the horn like structure of the experimental data occurs at different energy in comparison to the DPMJET III model for both $\frac{K^+}{\pi^+}$ and $\frac{K^-}{\pi^-}$ ratio.

Comparing table 1 and table 2 it may be said that significant differences between the values of $\frac{K^+}{\pi^+}$ and $\frac{K^-}{\pi^-}$ exist for both the models up to $\sqrt{s}$ =200GeV. However, at the higher energy regime ($\sqrt{s}$ =900GeV -7 TeV) no significant difference occurs between the values of $\frac{K^+}{\pi^+}$ and $\frac{K^-}{\pi^-}$ for both UrQMD and DPMJET III model. In case of the experimental data it can be seen that from $\sqrt{s}$ =200GeV the difference between the values of $\frac{K^+}{\pi^+}$ and $\frac{K^-}{\pi^-}$ is insignificant. The difference between the $\frac{K^+}{\pi^+}$ and $\frac{K^-}{\pi^-}$ values can be explained from the underlying physics of kaon and anti-kaon production mechanism. Here it should be mentioned that there are two possible mechanisms of kaon production, the associated production mechanism and the pair production mechanism. According to the associated production mechanism only $K^+$ mesons are produced by the following two interactions: $N+N \rightarrow N+X+K^+$ and $\pi + N \rightarrow X + K^+$. Where $N$ is the nucleon and X is either $\Lambda$ hyperons or $\Xi$ hyperons. On the other hand pair production mechanism produces $K^+$ and $K^-$ according to the interaction given by $N+N \rightarrow N+N+K^+ +K^-$. At the lower energy, the associated production mechanism dominates. As the energy increases, the pair production, which produces the same number of $K^+$ and $K^-$ becomes more significant. At higher energy the anti kaon excitation function is steeper than that of the kaon because of a higher threshold. So at higher energy the anti kaon production cross section increases faster than that of kaon and the $\frac{K^-}{\pi^-}$ ratio increases.

## 3.3 Energy Dependence Studies of Total kaon to pion multiplicity ratio

We have also calculated the total kaon to pion multiplicity ratio $\left(\frac{K^+ + K^-}{\pi^+ + \pi^-} = \frac{K}{\pi}\right)$ at these different collision energies for the proton-proton collisions and presented the values in table 3 for both UrQMD and DPMJET III simulated events. It can be noticed from the table that as energy increases, the $\frac{K}{\pi}$ ratio increases initially for both UrQMD and DPMJET III model. At higher energy in the LHC range the kaon to pion ratio becomes almost independent of energy in case of UrQMD model. But for DPMJET III simulation the values of $\frac{K}{\pi}$ go on increasing slowly with energy. No clear energy independency is observed for DPMJET III model in LHC energy regime. Moreover, the observed sudden decrease of $\frac{K^+}{\pi^+}$ and $\frac{K^-}{\pi^-}$ values vanishes in case of total kaon to pion multiplicity ratio $\left(\frac{K^+ + K^-}{\pi^+ + \pi^-} = \frac{K}{\pi}\right)$ in DPMJET III simulation. From table 3 it can be noted that the UrQMD simulated values of kaon to pion ratio are higher than the DPMJET III simulated values as energy increases from $\sqrt{s} = 200$ GeV. At energy less than 200 GeV, however, DPMJET III model calculated values of kaon to pion ratio are higher in comparison to the UrQMD simulated values.

Experimental studies of total pion to kaon multiplicity ratio $\left(\frac{K^+ + K^-}{\pi^+ + \pi^-} = \frac{K}{\pi}\right)$ in pp collisions have been reported by different collaborators in different times over a wide range of energy. From the report of the NA61 collaboration [21] we have calculated the values of $\frac{K}{\pi}$ ratio in pp collisions at $\sqrt{s} =$ 6.3GeV and presented the value in table 3 along with the values obtained from our simulated analysis. From the study of NA49 collaboration [16, 22-23] on pp collisions at $\sqrt{s}$ =17.3 GeV, we have calculated the values of $\frac{K}{\pi}$ ratio. In the regime of RHIC data, we extracted the values of $\frac{K}{\pi}$ ratio from the analysis of PHENIX collaboration [17] at $\sqrt{s}$ =62.4GeV and from the analysis of STAR collaboration at $\sqrt{s}$ =200GeV [18]. At $\sqrt{s}$=900 GeV and $\sqrt{s}$ =2.76TeV, the $\frac{K}{\pi}$ ratio have been calculated from the study of ALICE collaboration [24]. ALICE Collaboration [20] studied the pion, kaon and proton production in pp collisions at $\sqrt{s}$ =7 TeV also. In that paper they calculated the values of $\frac{K}{\pi}$ ratio in pp collisions. They have mentioned the values of $\frac{K}{\pi}$ ratio in pp collisions at different energy studied earlier and presented a study of energy dependence. In ref [24] the values of $\frac{K}{\pi}$ ratio at $\sqrt{s}$ =200GeV and $\sqrt{s}$=900 GeV have been mentioned in the text with proper references. The experimental values of $\frac{K}{\pi}$ ratio at different energies have been calculated from the plot given in [19] with the help of a high accuracy digitized graphical software as mentioned earlier.

Experimentally calculated values of $\frac{K}{\pi}$ ratio in pp collisions at $\sqrt{s} = 6.3, 17.3, 62.4, 200$ and 900 GeV and also at $\sqrt{s}$ =2.76 TeV and 7 TeV have been taken from these literatures and presented in table 3. From table 3 it can be seen that the experimentally obtained values of $\frac{K}{\pi}$ ratio increases initially with the increase of energy up to 200GeV. At energy greater than 200 GeV the $\frac{K}{\pi}$ ratio becomes independent of energy. Moreover, table 3 reflects that the experimentally obtained total kaon to pion multiplicity ratio agrees well with their UrQMD counterpart qualitatively and quantitatively over the entire energy range.

In comparison to the UrQMD analysis, DPMJET III model calculated values of $\frac{K}{\pi}$ ratio are higher than the experimental data up to 200GeV. As we enter in the LHC region (900 GeV to 7 TeV) DPMJET III simulated values of $\frac{K}{\pi}$ ratio are found to be lower than the corresponding experimental values. We have also studied the variation of $\frac{K}{\pi}$ ratio with energy graphically for the experimental, UrQMD simulated events and DPMJET III simulated events. Figure 3 depicts the variation of kaon to pion ratio with energy in case of pp collisions from 6.3 GeV to 7 TeV for experimental, UrQMD simulated and DPMJET III simulated events. From figure 3 it can be noticed that no horn like structure is observed for the experimental data when the energy dependence of total kaon to pion multiplicity ratio is studied.

It may be mentioned here that Hai-Yan Long et al [25] utilized the parton and hadron cascade model PACIAE based on PYTHIA to investigate the kaon to pion ratio in pp collisions at RHIC and LHC energy. They found that the PACIAE model calculated values of $\frac{K}{\pi}$ at $\sqrt{s}$ =17.2, 200 and 900 GeV agree with the NA49 [16, 22-23], STAR [18] and ALICE data [24, 26]. With the inclusion of the results for $\sqrt{s}$ =2.36, 7 and 14 TeV, it was found that the $\frac{K}{\pi}$ ratio increase slightly from $\sqrt{s}$ =0.2 to 0.9TeV and then saturates. Our study with UrQMD model predicts the same result. It should be mentioned here that ALICE collaboration also in their published papers [24, 26] studied the energy dependence of $\frac{K}{\pi}$ ratio in pp collisions.

## 4. Conclusions and Outlook

To summarize we recall that we have presented a systematic study of $\frac{K^+}{\pi^+}$, $\frac{K^-}{\pi^-}$ and $\left(\frac{K^++K^-}{\pi^++\pi^-}=\frac{K}{\pi}\right)$ ratio in proton-proton collisions as a function of the bombarding energy from 6.3 GeV to 7 TeV using UrQMD model and DPMJET III model. Comparisons of the simulated results with the available experimental data have also been presented. Important findings of this analysis are given below.

(1) Values of $\frac{K^+}{\pi^+}$ and $\frac{K^-}{\pi^-}$ differ from each other in the lower energy regime ($\sqrt{s}$ =6.3 GeV-$\sqrt{s}$ =200GeV) for both UrQMD and DPMJET III model simulation. The difference becomes insignificant in the LHC energy range ($\sqrt{s}$=900GeV-$\sqrt{s}$ =7 TeV). Experimental study also supports this observation. This observation can be explained on the basis of Kaon production mechanism.

(2) UrQMD simulated values of $\frac{K^+}{\pi^+}$ and $\frac{K^-}{\pi^-}$ are distinctively different from those of DPMJET III simulated values over the entire energy range.

(3) The dependence of $\frac{K^+}{\pi^+}$ and $\frac{K^-}{\pi^-}$ with energy shows a significant difference between the two approaches of UrQMD and DPMJET III simulation.

(4) In case of UrQMD model the values of $\frac{K^+}{\pi^+}$, $\frac{K^-}{\pi^-}$ and $\left(\frac{K^++K^-}{\pi^++\pi^-}=\frac{K}{\pi}\right)$ increase with energy initially and then saturate in the LHC energy regime. DPMJET III simulated ratio of $\frac{K^+}{\pi^+}$, $\frac{K^-}{\pi^-}$ and $\left(\frac{K^++K^-}{\pi^++\pi^-}=\frac{K}{\pi}\right)$ do not show any saturation at LHC region.

(5) A horn like structure is observed in case of DPMJET III simulation during the variation of $\frac{K^+}{\pi^+}$ and $\frac{K^-}{\pi^-}$ with energy. The horn like structure is found to be wiped out when the total kaon to pion multiplicity ratio $\left(\frac{K^++K^-}{\pi^++\pi^-}=\frac{K}{\pi}\right)$ was considered. No horn like structure has been observed in case of UrQMD simulation.

(6) Comparison of our results with the experimental data of $\frac{K^+}{\pi^+}$, $\frac{K^-}{\pi^-}$ and total multiplicity ratio$\left(\frac{K^++K^-}{\pi^++\pi^-}=\frac{K}{\pi}\right)$ has also been presented whenever available. Experimental study of energy dependence of $\frac{K^+}{\pi^+}$ and $\frac{K^-}{\pi^-}$ shows the presence of horn like structure. No horn like structure is observed in case of energy dependence of total kaon to pion multiplicity ratio $\left(\frac{K^++K^-}{\pi^++\pi^-} = \frac{K}{\pi}\right)$ for the experimental data.

(7) The experimental data was found to exhibit energy dependence at the lower energy regime but the values of $\frac{K^+}{\pi^+}$, $\frac{K^-}{\pi^-}$ and the $\frac{K}{\pi}$ ratio become independent of energy as energy goes to the LHC range (900 GeV to 7 TeV). We have demonstrated that the experimentally obtained values of kaon to pion total multiplicity ratio ($\frac{K}{\pi}$ values) are well reproduced by the UrQMD model. DPMJET III model simulated values of $\frac{K}{\pi}$ ratio are little different from the experimental values of $\frac{K}{\pi}$ ratio.

Before we come to the end of our paper, let us discuss some important points. It has already been mentioned in the introduction that pp collisions provide a baseline for studying nucleus-nucleus collisions. In this regard the study is interesting and the main focus of the paper is on proton-proton collisions only. We are not interested to study the nucleus-nucleus collisions in the present paper. We have applied two popular models to study the energy dependence of $\frac{K^+}{\pi^+}$, $\frac{K^-}{\pi^-}$ and total kaon to pion multiplicity ratio$\left(\frac{K^++K^-}{\pi^++\pi^-} = \frac{K}{\pi}\right)$. The choice of models has been made such that one is the string interaction model and the other is the hadronic transport model. People have successfully applied the UrQMD model in simulating proton-proton collisions earlier [27-30]. These papers establish the credibility of applying the UrQMD model in simulating pp collisions. Regarding the application of DPMJET III model in pp collisions it may be mentioned that F.W. Bopp et al [31] studied the anti-baryon to baryon ratio in pp collisions using the DPMJET III model. In another paper Bopp et al [32] compared the DPMJET III model with pp collisions data at $\sqrt{s}$ =200GeV from PHOBOS collaboration for the pseudo-rapidity distribution of charged hadrons. Very recently we have studied the energy dependence of $\frac{K^-}{K^+}$ ratio in pp collisions at $\sqrt{s}$ =6.3-200GeV successfully using the DPMJET III model [33]. These studies [31-33] establish the feasibility of applying DPMJET III model in pp collisions.

Another important point in this study is the selection of energy range. The range of energy considered in this study is wide enough covering almost the total energy range available so far including SPS, RHIC and LHC energy. Here lies the importance of this work. There are very few works available in the literature which covers such a wide range of energy in studying the $\frac{K^+}{\pi^+}$, $\frac{K^-}{\pi^-}$ and $\frac{K}{\pi}$ ratio in pp collisions.

By studying the energy dependence of $\frac{K^+}{\pi^+}$, $\frac{K^-}{\pi^-}$ and $\frac{K}{\pi}$ by UrQMD and DPMJET III model and comparing the results with the experimental data, we want to communicate that the approach of UrQMD and DPMJET III model is significantly different in explaining the energy dependence of $\frac{K^+}{\pi^+}$, $\frac{K^-}{\pi^-}$ and total kaon to pion multiplicity ratio $\left(\frac{K^++K^-}{\pi^++\pi^-}=\frac{K}{\pi}\right)$. Our study allows to discriminate between the two models which are based on two different approaches. The behavior of UrQMD and DPMJET III model in explaining the energy dependence of $\frac{K^+}{\pi^+}$, $\frac{K^-}{\pi^-}$ and total kaon to pion multiplicity ratio $\left(\frac{K^++K^-}{\pi^++\pi^-}=\frac{K}{\pi}\right)$ in pp collisions and the comparison of experimental results is the main song of this paper.

**Acknowledgement**

The corresponding author Dr. S. Bhattacharyya acknowledges Prof. Nestor Armesto of Universidade de Santiago de Compostela, Spain for helping in using the DPMJET III code. Dr. Bhattacharyya also acknowledges Prof. Dipak Ghosh, Department of Physics, Jadavpur University and Prof. Argha Deb Department of Physics, Jadavpur University, for their inspiration in the preparation of this manuscript.

**Table 1**

| Energy $\sqrt{s}$ GeV | UrQMD simulated value of $\frac{K^+}{\pi^+}$ ratio | DPMJET III model simulated value of $\frac{K^+}{\pi^+}$ ratio | Experimental Values of $\frac{K^+}{\pi^+}$ ratio | Reference for the Experimental Values of $\frac{K^+}{\pi^+}$ ratio |
|---|---|---|---|---|
| 6.3 | .049±.001 | .065±.002 | .081±.002 | [15] |
| 17.3 | .083±.003 | .095±.003 | .109±.003 | [16] |
| 62.4 | .101±.002 | .105±.002 | .097±.002 | [17] |
| 200 | .110±.008 | .099±.005 | .104±.008 | [18] |
| 900 | .122±.004 | .115±.006 | ......... | ......... |
| 2760 | .124±.004 | .116±.006 | ........... | ................ |
| 7000 | .124±.005 | .120±.007 | .126±.006 | [19] |

Table 1 represents the values of $\frac{K^+}{\pi^+}$ ratio in pp collisions at $\sqrt{s} = 6.3$ GeV to 7 TeV in the framework of UrQMD ,DPMJET III and the experimental values.

**Table 2**

| Energy $\sqrt{s}$ GeV | UrQMD simulated value of $\frac{K^-}{\pi^-}$ ratio | DPMJET III model simulated value of $\frac{K^-}{\pi^-}$ ratio | Experimental Values of $\frac{K^-}{\pi^-}$ ratio | Reference for the Experimental Values of $\frac{K^-}{\pi^-}$ ratio |
|---|---|---|---|---|
| 6.3 | .026±.001 | .031±.002 | .038±.001 | [15] |
| 17.3 | .066±.003 | .077±.003 | .092±.003 | [16] |
| 62.4 | .093±.004 | .097±.002 | .081±.005 | [17] |
| 200 | .105±.005 | .116±.005 | .102±.005 | [18] |
| 900 | .120±.007 | .113±.006 | .121±.013 | [20] |
| 2760 | .123±.005 | .115±.006 | ............ | ................ |
| 7000 | .124±.006 | .119±.007 | .128±.004 | [19] |

Table 2 represents the values of $\frac{K^-}{\pi^-}$ ratio in pp collisions at $\sqrt{s} = 6.3$ GeV to 7 TeV in the framework of UrQMD, DPMJET III and the experimental values.

**Table 3**

| Energy $\sqrt{s}$ GeV | Experimental value of $\frac{K}{\pi}$ ratio | UrQMD model simulated Value of $\frac{K}{\pi}$ ratio | DPMJET III model simulated value of $\frac{K}{\pi}$ ratio | Reference for the experimental value |
|---|---|---|---|---|
| 6.3 | .045±.001 | .046±.001 | .052±.002 | [21] |
| 17.3 | .082±.003 | .075±.003 | .087±.003 | [16,22-23] |
| 62.4 | .094±.002 | .097±.004 | .101±.002 | [17] |
| 200 | .103±.008 | .108±.005 | .109±.005 | [18,24] |
| 900 | .123±.004 | .121±.007 | .114±.006 | [24] |
| 2760 | .124±.004 | .123±.005 | .116±.006 | [19] |
| 7000 | .124±.005 | .124±.006 | .119±.007 | [19] |

Table 3 represents the values of total kaon to pion multiplicity ratio $\left(\frac{K^+ + K^-}{\pi^+ + \pi^-} = \frac{K}{\pi}\right)$ in pp collisions at $\sqrt{s} = 6.3$ GeV to 7 TeV in the framework of UrQMD and DPMJET III model along with the experimentally obtained values.

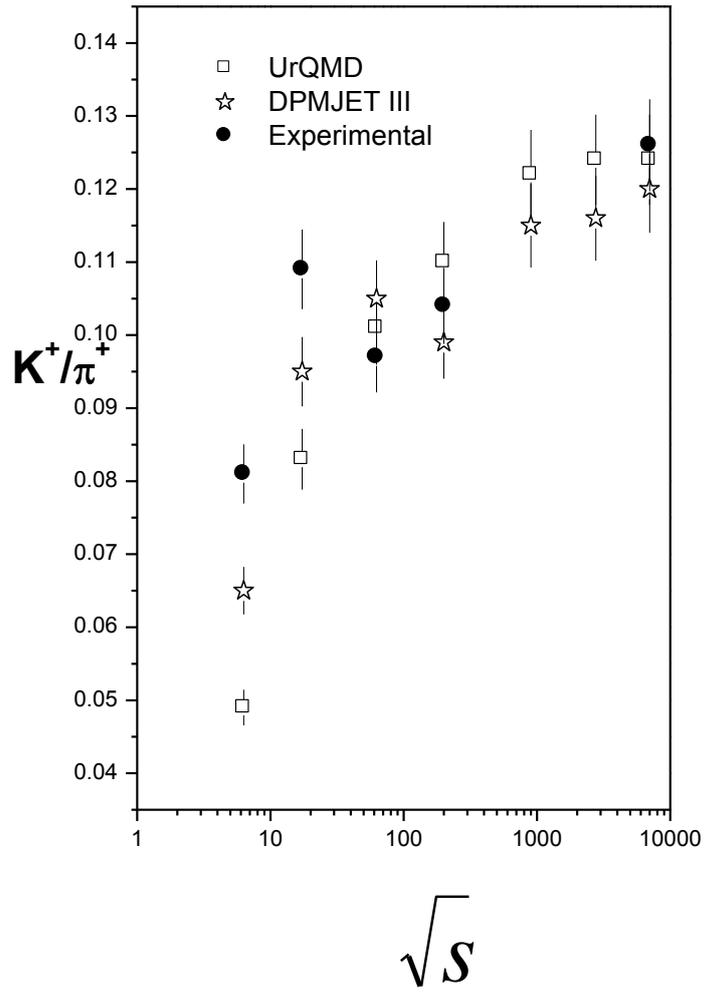

Fig 1 represents the energy dependence of $\frac{K^+}{\pi^+}$ ratio in pp collisions at $\sqrt{s} = 6.3$ GeV to 7 TeV of the UrQMD, DPMJET III and experimental analysis.

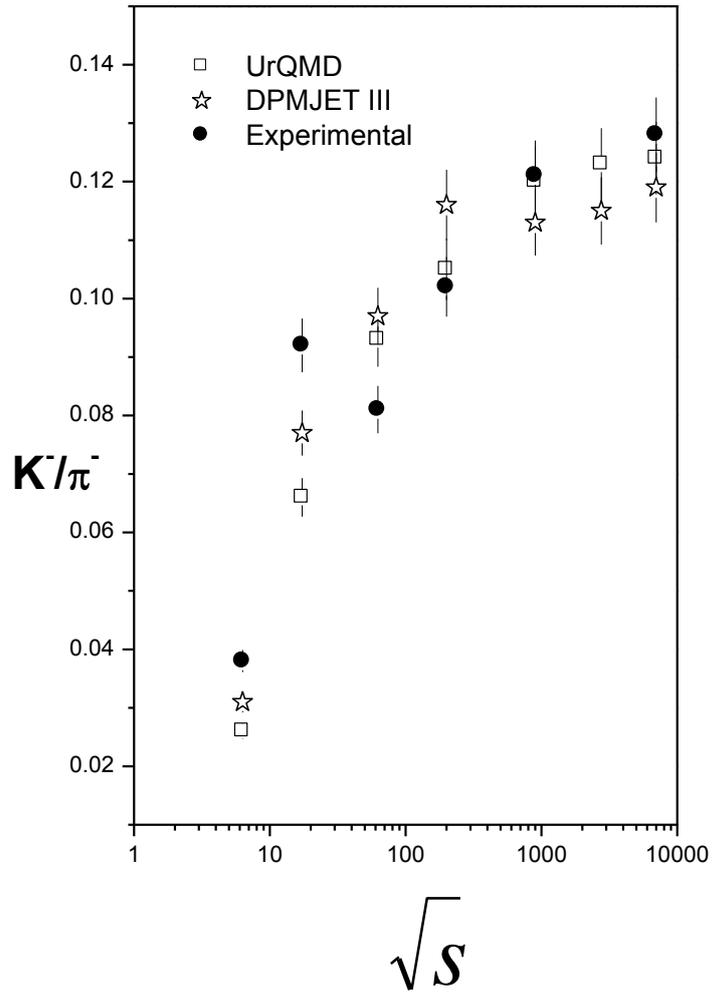

Fig 2 represents the energy dependence of $\frac{K^-}{\pi^-}$ ratio in pp collisions at $\sqrt{s}=6.3\,\text{GeV}$ to 7 TeV of the UrQMD, DPMJET III and experimental analysis.

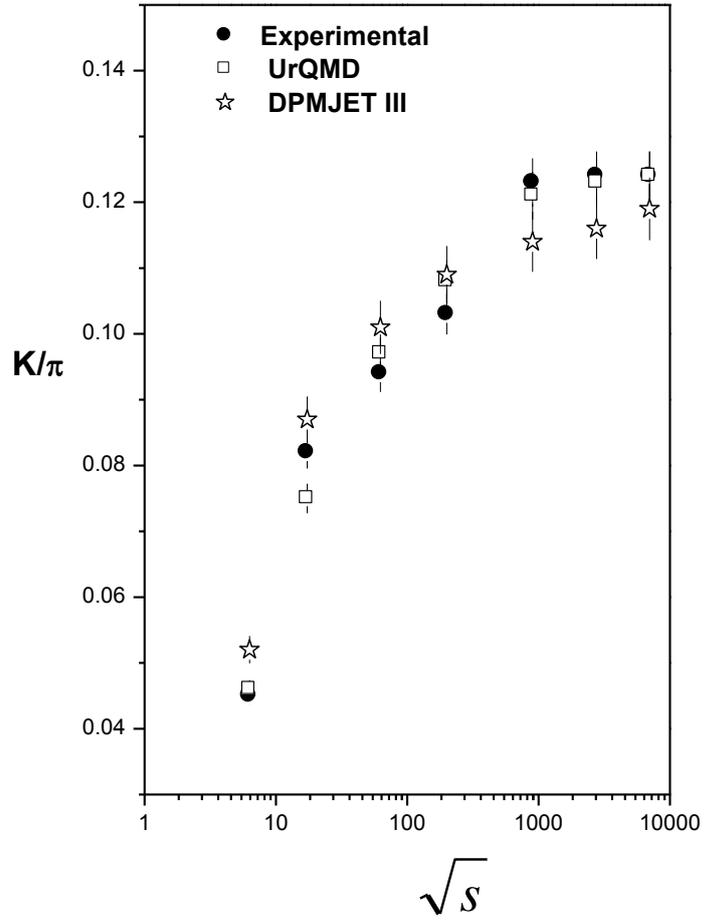

Fig3: Variation of total kaon to pion multiplicity ratio $\left(\frac{K^+ + K^-}{\pi^+ + \pi^-} = \frac{K}{\pi}\right)$ with energy $\sqrt{s}$ in pp collisions at $\sqrt{s} = 6.3\,\text{GeV}$ to 7TeV for the experimental data extracted from the papers of different collaborators, UrQMD model simulated values and DPMJET III model simulated values.

**There is no conflict of interest in publishing the paper**